\documentclass[twocolumn]{IEEEtran} %!PN

\def\BibTeX{{\rm B\kern-.05em{\sc i\kern-.025em b}\kern-.08em
    T\kern-.1667em\lower.7ex\hbox{E}\kern-.125emX}}

\def\ra{\rangle}
\def\la{\langle}
\def\rl{\rangle \langle}
\def\openone{\leavevmode\hbox{\small1\kern-3.8pt\normalsize1}}
\def\RR{{\rm I\kern-.2emR}}
\def\tr{{\rm tr}\; }
\def\ce{{\cal E}}
\def\cg{{\cal G}}
\def\cd{{\cal D}}
\def\cb{{\cal B}}
\def\cn{{\cal N}}

\def\cf{{\cal F}}
\def\ci{{\cal I}}
\def\ca{{\cal A}}
\def\cv{{\cal V}}
\def\cc{{\cal C}}

\def\on{^{\otimes n}}
\def\pn{^{(n)}}

\def\psirq{\psi^{RQ}}

\newcommand{\beq}{\begin{equation}}
\newcommand{\eeq}{\end{equation}}
\newcommand{\beqa}{\begin{eqnarray}}
\newcommand{\eeqa}{\end{eqnarray}}
\newtheorem{definition}{Definition}
\newtheorem{theorem}{Theorem}

\newtheorem{lemma}{Lemma}
\newtheorem{corollary}{Corollary}

\begin{document}
\title{On Quantum Fidelities and Channel Capacities}

\author{Howard~Barnum, \thanks{H. Barnum is with the 
School of Natural Science and ISIS, Hampshire College, Amherst, MA 01002.  E-mail:  {\tt hbarnum@hampshire.edu}.}
E.~Knill, \thanks{E. Knill is with the Los Alamos National Laboratories,
Mail Stop B265, 
Los Alamos, NM 87545.  E-mail: {\tt knill@lanl.gov}.} 
M.~A.~Nielsen \thanks{M. Nielsen is with the 
Physics Department, MC 12-33, California Institute of Technology, 
Pasadena CA 91125.
E-mail: {\tt mnielsen@theory.caltech.edu}.}}

\date{\today}

\markboth{Submitted to IEEE Transactions on Information Theory}
%\markboth{Draft for submission to IEEE Transactions}
{Barnum, Knill, and Nielsen: On Quantum Fidelities and Channel Capacities}
\maketitle

\begin{abstract}  
We show the equivalence of two different notions
of quantum channel capacity:
that which uses the entanglement fidelity as its criterion for
success in transmission, and that which uses the minimum fidelity
of pure states in a subspace of the input Hilbert space as its criterion.
As a corollary, {\em any} source with entropy less than the capacity may be
transmitted with high entanglement fidelity.  We also show that a restricted
class of encodings is sufficient to transmit any quantum source which may be
transmitted on a given channel.  This enables us to 
simplify a known upper bound for the channel capacity.  It also enables us
to show that the availability of an auxiliary classical channel from encoder
to decoder does not increase the quantum capacity.
\end{abstract}


\section{Introduction}
A theory of quantum information is emerging which shows striking parallels 
with,
but also fascinating differences from, classical information theory.  
One of the principal concerns of such theories is the
capacity of a noisy channel for transmitting the state of a system despite 
some uncertainty about that state; that is, 
for rendering the state of some other system virtually identical to the
initial state of
the system at hand.  In classical information theory, this is one of a set 
of mutually exclusive classical 
states; in quantum mechanics, a quantum state represented by a vector in a 
Hilbert space, or a density operator on that space.  Classically, the input
system may retain its original state, while the no-cloning theorem and 
related results
\cite{Wootters82a}, \cite{Dieks82a}, \cite{Barnum96a},\cite{Buzek96a}, 
\cite{Buzek97a}, \cite{Buzek98a},\cite{Bruss97a},
\cite{Niu98a},\cite{Cerf98a},\cite{Cerf98b}, \cite{Werner98a},\cite{Keyl98a}
imply that in the quantum case the input system cannot in general
remain in its initial state.  Both theories allow the use
of encoding and decoding operations to increase the fidelity with which
states are transmitted.  Due partly 
to the peculiarly quantum fact that a system's state may be entangled with
that of other systems, a greater variety of definitions of capacity has arisen
in quantum mechanics, depending, for example, on whether the entanglement of
a system with some reference system is required to be preserved by the 
transmission process, or not.  Here we concentrate on two 
notions of quantum capacity, 
one investigated for example in 
\cite{Schumacher96a},\cite{Schumacher96b},\cite{Nielsen96c},\cite{Barnum97a}, 
concerned with the maximum entropy of a density operator 
whose entanglement with a reference system which does not undergo the noise
process can be preserved with high fidelity, and another arising for example
in 
\cite{Bennett96a},\cite{Lloyd97a},
\cite{Bennett97a},\cite{DiVincenzo97a} and concerned with the maximum
size of a Hilbert
space all of whose pure states can be preserved 
with high fidelity.
We show that these two 
definitions of capacity are in fact equivalent, in the situation in which 
sources are required to satisfy the quantum analogue of the asymptotic
equipartition principle.  We also show that any source with 
entropy less than the capacity may be sent with high entanglement fidelity, so
that quantum entropy and capacity parallel classical entropy and capacity in 
this respect.

We also establish that any source that may be transmitted may be transmitted
using only a maximal partial isometry as an encoding.  This can be interpreted 
as meaning that encoding
can be a unitary process, except for an initial projection of the 
source onto
a subspace small enough to fit into the channel, if the channel is smaller
than the source.  

%(This procedure may
%be trace-decreasing, hence not physical, but {\em any} extension of the 
%partial isometry to a trace-preserving encoding is usable as an encoding.
%The interpretation is just that it doesn't matter what we do if our initial
%measurement of the projector onto the subspace small enough to fit into the
%channel, and the complementary projector, has the result corresponding to 
%the complementary projector.) 

This fact, which is in some ways analogous to the source-channel coding 
separation theorem of classical information theory, allows us to simplify
a known upper bound on the quantum channel capacity, by removing from the
expression a maximization over encodings, confirming an earlier conjecture.
The conjecture has also been confirmed by \cite{Barnum97b}, but the result that
{\em any} source that may be transmitted may be transmitted using partially
isometric encodings is slightly stronger 
than that obtained in \cite{Barnum97b}.

	In \cite{Adami97a} Adami and Cerf express the view
that ``Whether a
capacity can be defined {\em consistently} that characterizes the
``purely'' quantum component of a channel is an open question.''  In
our view, the pure-state capacity defined below and in earlier papers
is just such a consistently defined capacity, and the result that any
source with entropy less than the entanglement capacity of a channel
may be transmitted with high entanglement fidelity removes the last
possible objection to the capacity for entanglement transmission as
another such notion of ``purely quantum'' capacity.

We note that besides those cited above, 
many authors have worked on the problem of quantum 
information transmission through quantum channels; some of this work
calculates or 
or bounds the capacity we study here, for particular channels or classes of
channels:  an incomplete list that could serve as an entry to the literature 
includes \cite{Bennett96a}, \cite{Shor96b}, \cite{DiVincenzo97a}, 
\cite{Lloyd97a}, \cite{Bennett97a}, 
\cite{Bruss97a}, \cite{Cerf98a}, \cite{Cerf98c}. 
Some of the extensive literature on the more algebraic approach to quantum
coding also yields information about the quantum capacity.

\section{Quantum sources and channel capacity}
\subsection{Mathematical preliminaries and notation}
\label{section: preliminaries}
The effect of 
encoding procedures, decoding procedures, 
and noisy quantum channels
on the state of a system may be described by 
 completely positive linear maps 
$\cn$, from the space $B(H_c)$ of bounded linear operators
on a input Hilbert space $H_c$, to the space $B(H_o)$
of bounded linear operators
on an output Hilbert space $H_o$
\cite{Hellwig70a},\cite{Kraus83a},\cite{Choi75a}.  In this paper, we
consider only discrete channels, which we define as having finite-dimensional
input and output Hilbert spaces (the word ``bounded'' in the specification
of the input and output spaces 
is redundant in the discrete case).  We will sometimes use the
term {\em quantum operation} for a trace-nonincreasing completely
positive map.  Such maps have representations in terms of linear
operators $A_i$ \cite{Hellwig70a},\cite{Kraus83a},
\beqa
\ca(\rho) = \sum_i A_i \rho A_i^\dagger\;,
\eeqa
with
\beqa 
\sum _i A_i^\dagger A_i \le I\; ;
\eeqa
equality holds in the latter when the map is {\em trace-preserving}.
We call the set $\{A_i\}$ an {\em operator decomposition}, or simply
decomposition, of the operation $\ca$, and sometimes write:
\beqa
\ca \sim \{A_i\}
\eeqa
to indicate that $\{A_i\}$ is an operator decomposition of $\ca$.
Any two decompositions of the same operation,   
$\{A_i\}$ having $r$ operators
and $\{B_i\}$ having $s \le r$ operators, 
are related by \cite{Choi75a}:
\beqa \label{eq: unitary remixing}
A_i = \sum_{j=1}^s m_{ij} B_j
\eeqa
where $m$ is the matrix of a maximal partial isometry from the complex
vector space $\cc^s$ to 
$\cc^r$.
A partial isometry is a generalization of a unitary 
operator, which must satisfy $V V^\dagger = \Pi$ for some projector
$\Pi$.  Such an isometry will then also satisfy $V^\dagger V = \Gamma$
for some projector $\Gamma$ having the same dimensionality as $\Pi.$
If the range and domain spaces
of a linear operator $V$ 
have different dimensionality, it will not be possible to
find a unitary mapping between the two:  the best one can do is find a
partial isometry $V$ such that one of $V  V^\dagger$ and $V^\dagger V$ is 
the identity (whichever one operates on the smaller space).    
We 
will call such a map a {\em maximal} partial isometry between
the spaces $S_1$ and $S_2$.  
A partial isometry with $ V V^\dagger$ having dimension
$C$ may be thought of as projecting onto a $C$-dimensional 
subspace of $V$'s domain
Hilbert space and then mapping that subspace
unitarily to a $C$-dimensional subspace of the range Hilbert
space.  
Thus if $s \le r$ in \ref{eq: unitary remixing}, 
$m$'s columns are 
$s$ orthonormal vectors in $\cc^r$:
\beqa
\sum_j m_{ij}^* m_{kj} = \delta_{ik}\;
\eeqa
or in other words:
\beqa
m m^\dagger  = I^{(s)}\;.
\eeqa

Sometimes an operation $\ca$ will have a decomposition consisting of a
single operator $A$;  in this case, we will often use the roman
letter  $A$ 
to denote the operation $\ca$ as well as the operator
$A$ when no confusion will result.   We note
that care is needed when the operator includes a scalar factor $z$:  
thus if $\ca \sim \{A\}$ while $\cb \sim \{zA\}$, 
we may also 
refer to the operation $B$ as either $zA$ or $|z|^2\ca$.

We write
$\ca \ce$ for the operation of $\ce$ followed by $\ca$;  thus
$\ca\ce(\rho) \equiv \ca(\ce(\rho))$.

Any quantum operation on a system $Q$ 
may be realized \cite{Stinespring55a},\cite{Hellwig70a},\cite{Kraus83a} 
by a ``unitary representation''
in which the Hilbert space $Q$ is extended by adjoining an environment
$E$ prepared in a standard state $|
0^E\rangle$, and the system and environment undergo a unitary interaction,
followed by a projection on the environment system.  Any such unitary
interaction with a given initial environment state 
determines a quantum operation.  (In the case of a 
trace-preserving operation, the environment projection is the identity.)
That is,
\beqa 
\ca(\rho) = \tr_E (\pi^{E} U^{QE}|0^E\ra\la 0^E| \otimes \rho^{Q} 
U^{\dagger QE} \pi^{E})\;.
\eeqa
The operators $A_i$ in the operator decomposition representation discussed
above, turn out to be the ``operator matrix elements''
\beqa
A^Q_i = \la i^E| U^{QE} |0^E\ra
\eeqa
of the unitary interaction, between the initial 
environment state and orthonormal
environment vectors 
vectors $|i\ra$ of the basis used for the partial trace over the environment.
The freedom (\ref{eq: unitary remixing}) 
to ``unitarily remix'' the operators $A_i$, 
obtaining another valid decomposition, 
is just the freedom to do the enviroment partial trace in a 
different environment basis (related to the first by that same unitary).

\subsection{Transmission and capacity}
We now review the problem of entanglement 
transmission, as discussed more fully in \cite{Schumacher96a},
\cite{Schumacher96b},\cite{Barnum97a}.
A fuller discussion of the problem may be found in those articles.
Here the goal is to use block coding to send the density operator
of a source in a manner which preserves its entanglement with whatever
reference system it may be entangled with.  We imagine the density operator
$\rho^Q$ of our quantum system to arise from a pure state on a larger composite
system $RQ$, by tracing out the ``reference'' system $R$.  That is,
\beqa
\rho^Q = \tr_R {(|\psi^{RQ}\ra\la\psirq|)}\;.
\eeqa
For $R$ with dimension at least as great as that of $\rho^Q$'s support, 
such purifications always exist;  different purifications
of the same $\rho^Q$ are related by unitary transformations on $R$.
We define
the entanglement fidelity as 
\beqa
F_e(\rho^Q, \ca) \equiv
\langle \psirq | \ci \otimes \ca(|\psirq \rangle \langle \psirq|)|
\psirq \rangle \;, 
\eeqa
the matrix element of the final, noise-affected state of the system $RQ$, with the initial 
state $|\psi^{RQ}\ra$.  This is easily shown to be independent of which 
purification $|\psirq\ra$ is used, and to have the form:
\beqa
F_e(\rho, \ca) = \sum_i |\tr A_i \rho|^2.
\eeqa
Note that while \cite{Barnum97a} defined $F_e$ as 
the renormalized entanglement
fidelity $\sum_i |\tr A_i \rho|^2/\tr \ca(\rho),$ we have omitted the
normalization, since the unrenormalized version is most useful in the
present context.  When we need the renormalized entanglement fidelity,
just defined, we will use the symbol $\hat{F}_e$.

We define a {\em quantum source} $\Sigma = (H_s,\Upsilon)$ to consist of a
Hilbert space $H_s$ and a sequence $\Upsilon =
\{\rho_s^{(1)},\rho_s^{(2)},...,\rho_s\pn,...\}$ where $\rho_s^{(1)}$ is a density 
operator on $H_s$, $\rho_s^{(2)}$ a density operator on $H_s \otimes
H_s$, and $\rho_s\pn$ a density operator on $H_s^{\otimes n},$ etcetera.
We define the {\em entropy rate} 
of a source $\Sigma$ as 
\begin{eqnarray}
S(\Sigma) \equiv 
\limsup_{n \rightarrow \infty} \frac{ S(\rho_s\pn)}{n}.
\end{eqnarray}
(Sometimes we use the term ``entropy of a source'' to mean 
its entropy rate.)
A {\em quantum channel} will be a trace-preserving map 
\beq
\cn: B(H_c) \rightarrow B(H_o)\;
\eeq
from operators over a channel input space $H_c$ to operators over a channel output space $H_o$.
A {\em coding scheme} for a given 
source into a given channel consists of a sequence
$(\ce\pn,\cd\pn)$  of trace-preserving encoding maps 
and decoding maps 
\beqa
&\ce\pn&: \;\;B(H_s\on) \rightarrow B(H_c\on) \nonumber \\
&\cd\pn&: \;\;B(H_o\on) \rightarrow B(H_s\on) \;.
\eeqa
We say that a source $\Sigma$ may be sent reliably over a quantum channel
$\cn$ if there exists a coding scheme such that 
\beqa
\lim_{n \rightarrow \infty} F_e(\rho\pn, \cd\pn  \cn\on   \ce\pn)
= 1\;.
\eeqa
We say that rate $R$ is achievable with a quantum channel $\cn$ if there
is a source $\Sigma$ with entropy $R$
which may be sent reliably over the channel.  We define the quantum capacity
of the channel for transmission of entanglement, 
$Q_e(\cn),$ as the supremum of rates achievable with the channel $\cn$.
This definition of channel capacity leaves open the possibility that although 
some sources with entropy close to the capacity can be sent reliably, not
all such sources can.  Classically, it turns out that this is not the case:
any source with entropy less than the classical
capacity may be sent reliably.  In 
what follows, we will establish that this is also the case for the quantum
capacity.  We will also establish the equality of the capacity for 
entanglement transmission $Q_e,$ with the capacity for transmission of pure
states in a subspace, $Q_s,$ used for example in \cite{Bennett97a},
\cite{Lloyd97a}.  
We define the minimum pure-state fidelity, or simply pure-state fidelity,
of a subspace $H$ of the channel input Hilbert space as
\beqa
F_p(H,\ca) \equiv 
{\rm min}_{|\psi\rangle \in H}
\langle \psi | \ca (|\psi \rangle \langle \psi|)
|\psi \rangle \;.
\eeqa
We say the rate $R$ of transmission of subspace dimensions is achievable 
with channel $\cn$ if
there exists a sequence of subspaces $H\pn$ of $H_c\on$ such that
\beqa
\limsup_{n \rightarrow \infty} \frac{\log{{\rm dim}(H\pn)}}{n} = R
\eeqa
and there is a coding scheme which sends it reliably in the sense that
\beqa
\lim_{n \rightarrow \infty}  F_p(H\pn,\cd\pn  \cn\on  \ce\pn) = 1.
\eeqa 
We define the capacity of the channel $\cn$ 
for transmission of subspaces, $Q_s$, 
as the supremum of achievable rates of transmission of subspace dimensions 
with channel $\cn$.

\subsection{The Quantum Asymptotic Equipartition Property}

The {\em $\epsilon$-typical subspace} 
for an $n$-block of material $\rho\pn$ 
produced by a quantum source $\Sigma$ 
on a Hilbert space $H$ is defined to be
the subspace $T\pn_\epsilon$ of $H\on$ spanned by the eigenvectors 
$|\lambda\rangle$ of $\rho\pn$ whose
eigenvalues $\lambda$ satisfy:
\beqa \label{eqtn: typical eigenvalues}
2^{-n(S(\Sigma) +
              \epsilon)} \le \lambda \le 2^{-n(S(\Sigma) - \epsilon)}\;.
\eeqa
An equivalent requirement is:
\beqa
|-\frac{1}{n}\log{\lambda} - S(\Sigma)| \le \epsilon.
\eeqa

The definition derives its interest from the fact that for some
interesting sources---for example, the i.i.d. source with
$\rho\pn = \rho\on$ \cite{Schumacher95a}---all 
but a negligible portion of the source
becomes concentrated in an $\epsilon$-typical subspace as 
$n$ goes to infinity, no matter how small $\epsilon$ is chosen 
to be.  More formally, the i.i.d. source satisfies the {\em Quantum Asymptotic
Equipartition Property} (QAEP).
(Here and elsewhere, we will sometimes use the phrase ``for large enough
$n$, $P(n)$ is true'' to mean ``there exists an $n_0$ such that for 
all $n>n_0$, $P(n)$ is true''.)

\begin{definition}\label{definition: QAEP}
A source $\Sigma = \{\rho^{(1)},...,\rho\pn,...\}$ is said to satisfy the Quantum
Asymptotic Equipartition Property if 
for any positive $\epsilon$ and $\delta$,
for large enough $n$
the $\epsilon$-typical
subspace of $\rho\pn$ satisfies:
\beqa \label{eqtn: quantum AEP}
{\rm tr}\,\Lambda\pn \rho\pn \Lambda\pn > 1 -
              \delta\;,
\eeqa
where $\Lambda\pn$ is the projector onto $T\pn_\epsilon$.
\end{definition}

An immediate consequence of satisfaction of the QAEP is the
following bound on the the
dimension of the typical subspace, which holds for $n$ large
enough that the trace bound in the QAEP is satisfied:
\beqa
(1 - \delta) 2^{n(S(\Sigma) - \epsilon)} 
\le {\rm dim} (T\pn_\epsilon) \le
2^{n(S(\Sigma) + \epsilon)}\;,
\eeqa 

A slightly more involved consequence is that for large enough $n$ 
no subspace of dimension smaller than the lower
bound $(1 - \delta)2^{n(S(\Sigma) - \epsilon)}$ 
on the size of the typical subspace, has probability
greater than $\delta$.  That is, if $\Pi$ is the projector
onto such a space,
\beqa
\tr \Pi \rho \le \delta\;.
\eeqa
See \cite{Jozsa94a}.

The classical 
Shannon-McMillan-Breiman theorem states that all stationary ergodic
classical sources satisfy the (classical) AEP;  however, 
these are not necessarily all the sources which satisfy it.
There is as yet no known quantum analogue of the Shannon-McMillan-Breiman
theorem, providing a broad and natural class of sources satisfying 
the QAEP, although there has been work in this direction \cite{King98a}.

\section{Useful facts about fidelities}

\subsection{Convexity of Entanglement Fidelity in the Input Density Operator}
\begin{lemma}
The entanglement fidelity is convex in the input density operator,
\beqa 
F_e ( \lambda \rho_1 + (1 - \lambda) \rho_2) , \ce)
& \le &  \lambda F_e(\rho_1, \ce) + (1 - \lambda) F_e ( \rho_2, \ce)\;.
\nonumber \\
& & 
\eeqa
\end{lemma}
\begin{proof}
Note that the entanglement fidelity may be viewed as
the squared norm $||{\bf a}||^2 \equiv 
\sum_i |a_i|^2$ of a complex vector ${\bf a}$ whose components are:
\beqa
a_i \equiv \tr A_i \rho_1\;.
\eeqa
Then, letting also
\beqa
b_i \equiv \tr A_i \rho_2 \;,
\eeqa 
the entanglement fidelity of the 
convex combination of $\rho_1$ and $\rho_2$ may be written
\beqa
F_e ( \lambda \rho_1 + (1 - \lambda) \rho_2,\ce)
= ||\lambda {\bf a} + (1 - \lambda){\bf b}||^2 \;.
\eeqa
Any norm is easily shown to be convex 
(see e.g. \cite{Rockefellar70a} for real vector spaces), and since
a norm is positive its square is also convex
and the lemma follows.
\end{proof}

Note that with this representation of the entanglement fidelity, the
freedom to choose an environment basis (equivalently, the freedom to 
move to a different operator decomposition of a given operation) 
corresponds to performing a maximal partial isometry $V$ from the complex
vector space containing ${\bf a}$ to another complex vector space
(with dimension equal to the number of operators in the new decomposition).
(Since the transformation is length-preserving, 
it preserves (as it had better!)
the entanglement fidelity.)  We may use this ``unitary'' freedom to transform 
the vector ${\bf a}$ into one of the same length with only a particular component,
 say the first, nonzero.  Then the entanglement fidelity will just be the
modulus $|\tr A_1 \rho|^2$ of that component.  This gives us a useful 
lemma:

\begin{lemma}
\label{l:single}
There exists an operator sum decomposition $\{A_i\}$ of $\ca$
such that $F_e(\rho,\ca) = F_e(\rho, A_1)$.
\end{lemma}

It may be instructive to see how this result arises in the
$RQE$ or unitary view of operations.  The 
entanglement fidelity is the fidelity of 
$\rho^{RQ'}$ and the initial state of $RQ$;  this is equal to the 
squared 
inner product of $|0^E\rangle|\psirq\rangle$ with some purification of 
$\rho^{RQ'}$.  The final pure state of $RQE$ is such a purification,
so it is related to the one whose inner product with the initial
state gives the fidelity by a unitary on the environment;  view
this inner product as one between the final state of $RQE$ and 
some other tensor product state 
$U^{\dagger E} |0^E\rangle|\psirq\ra$, and the result
follows (since the individual terms in the entanglement fidelity
correspond to particular states in an orthonormal
basis used for the trace over the
environment). 

A slight variant of this interpretation is useful in the proof of
the next lemma.  Write the entanglement fidelity as
\beqa
& & \tr_{RQ} (\rho^{RQ'} |\psi^{RQ}\ra \la \psi^{RQ}|) \nonumber \\
&=& \tr_{RQE} (U^{QE} |0^E\ra|\psi^{RQ}\ra \
\la \psi^{RQ}|\la 0^E| U^{QE\dagger} \nonumber \\
&{}& \times (|\psi^{RQ}\ra \la \psi^{RQ}| \otimes I^E ))\nonumber \\
&=& 
||(|\psi^{RQ}\ra \la \psi^{RQ}| \otimes I^E)  
U^{QE} |\psi^{RQ}\ra |0^E\ra ||^2\;.
\eeqa
That is, the entanglement fidelity is just squared length of the
projection $|\pi^{RQE'}\rangle$ 
of the evolved pure state of $RQE$ onto the tensor 
product of the environment and the one dimensional subspace
of $RQ$ spanned by the initial state of $RQ$.  The vector 
${\bf a}$ above is in fact just this projection;  the components of
${\bf a}$ are the individual terms in the entanglement fidelity in 
a particular operator decomposition, i.e. the components of the 
vector ${\bf a}$ in a particular orthonormal basis.  These correspond to the
components of the projection   $|\pi^{RQE'}\rangle$ in a particular orthonormal
basis
$|\chi_i^E\rangle|\psi^{RQ}\rangle$
for the subspace onto which we have projected, which corresponds to a 
choice of orthonormal basis $|\chi_i^E\rangle$ for the environment.  So the 
lemma above is nothing but the observation that if we do the trace
 (in the definition of the entanglement fidelity) in an 
environment basis the first vector of which is a normalized version 
of $|\pi^{RQE'}\rangle,$ we only get one term, which is of course
the length of this projection.

We use this point of view to derive 
a lemma which concerns applying operations in sequence:  if an operation
has high fidelity, then the fidelity of the operation consisting 
of that operation 
followed by a second operation, is close to the fidelity of the 
second operation alone.

\begin{lemma} \label{lemma: composition}
If $F_e(\rho, \ce) \ge 1 - \eta$ then for trace-nonincreasing $\ca$, 
\beqa
|F_e(\rho, \ca  \ce) - F_e(\rho, \ca) | \le  2\eta\;.
\eeqa
\end{lemma}
\begin{proof}
Let $E1$ and $E2$ be environments inducing the operations
$\ce$ and $\ca$ through unitary interactions $U^{QE1}$ and
$V^{QE2}$ respectively.  Then:
\beqa
1 - \eta &\le& 
F_e(\rho, \ce) \nonumber \\
&=& || |\psirq\ra\la\psirq| \otimes I^{E1} U^{QE1} 
|\psirq\ra|0^{E1}\ra ||^2 \nonumber \\
& = &  \la \psirq | \la 0^{E2} | \la \chi^{E1} |U^{QE1} 
|0^{E1}\ra |0^{E2}\ra |\psirq\ra \nonumber \\
{}&{}&{} 
\eeqa
for some $|\chi^{E1}\ra$.  That is, the two vectors \newline
$U^{QE1}
|0^{E1}\ra |0^{E2}\ra |\psirq\ra$ and $|\chi^{E1}\ra |0^{E2}\ra
|\psirq\ra$ are close.
Now consider the two fidelities the magnitude of whose difference
we wish to bound;  these may be written
as the squared lengths of projections of the two close vectors just
considered.  That is, define
\beqa
P \equiv 
|\psirq\ra\la\psirq| \otimes I^{E1 E2} \;.
\eeqa
Then 
\beqa
&{}&F_e(\rho, \ca \ce)  \nonumber \\ 
&=&|| P
V^{QE2} U^{QE1}
|\psirq\ra|0^{E1}\ra|0^{E2}\ra ||^2 \nonumber \\
& = & || (V^{\dagger QE2} P V^{QE2} )
U^{QE1}
|\psirq\ra|0^{E1}\ra|0^{E2}\ra ||^2 \eeqa
and
\beqa
F_e(\rho, \ca ) & = & 
|| P V^{QE2} U^{QE1}
|\psirq\ra|0^{E1}\ra|0^{E2}\ra ||^2 \nonumber \\
& = &  || (V^{\dagger QE2} P V^{QE2} )
|\psirq\ra|\chi^{E1}\ra|0^{E2}\ra ||^2\;. \nonumber \\
&{}&
\eeqa
 
{}From elementary geometry, if for normalized $|1\ra$ and 
$|2\ra$,$~|\la 1 | 2 \ra|^2 = 1 - \eta$ then for any projector
$P$, $ |\la 1 | P |1 \ra - \la 2 | P | 2 \ra|  \le 2 \eta \;.$
This may be applied directly to obtain the lemma.  
\end{proof}

A very simple but useful lemma implies that if two operations have high 
entanglement fidelity on the same density operator, the final 
density operators have high fidelity with each other.
The notion of fidelity used here is treated in \cite{Uhlmann76a},
\cite{Jozsa94b}, and \cite{Bures69a}.  It may be defined by
\beq
F(\rho_1, \rho_2) \equiv \max_{|\psi_1\ra, \psi_2\ra} |\la \psi_1 | \psi_2 \ra|^2,
\eeq
where $|\psi_i\ra$ are purifications of $\rho_i$.

In terms of this fidelity, the lemma is:
\begin{lemma}\label{lemma: closefinal}
If $\ca$,$\cb$ are trace-preserving and 
$F_e(\rho, \ca) \ge 1 - \epsilon_1$ and $F_e(\rho,\cb) \ge 1 - \epsilon_2$
then $F(\ca(\rho),\cb(\rho)) \ge 1 - \epsilon_1 - \epsilon_2$.
\end{lemma}

\begin{proof}
Note that if $\la 1| 1\ra = \la 2 |2 \ra = 1$, $|\langle 1|2\rangle|^2
> 1 - \epsilon_1$ and $|\langle 1|3\rangle|^2
> 1 - \epsilon_2,$  
then $|\langle 2|3\rangle|^2 > 1 - \epsilon_1 - \epsilon_2$.
Apply this with $|2\rangle$ and $|3\rangle$ being the purifications of  
$\ca(\rho)$ and $\cb(\rho)$ whose squared inner products with a purification
$|1\rangle$ of $\rho$
give the entanglement fidelities, obtaining
\beqa
|\langle 2 | 3 \rangle |^2 
\ge  1 - \epsilon_1 - \epsilon_2 \;.
\eeqa
Since the fidelity is the maximum squared inner product of purifications,
$F(\ca(\rho),\cb(\rho)) \ge 
|\langle 2 | 3 \rangle |^2 \ge 1 - \epsilon_1 - \epsilon_2$, as claimed.
\end{proof}

\subsection{Continuity of entanglement fidelity in the input operator}
We will also need the continuity lemma
for entanglement fidelity, trivially extended to the case of 
unnormalized $F_e$ from  \cite{Barnum97a}.
\begin{lemma} \label{lemma: continuity of entanglement fidelity}
\begin{equation} \label{eqtn: unnormalized continuity lemma} 
|F_e(B + \Delta,{\cal A}) - F_e(B,{\cal A})| \le
({\rm tr}(|\Delta|))^2 + 2 {\rm tr}(|\Delta|),
\end{equation}
where $|\Delta| \equiv \sqrt{\Delta^\dagger \Delta}.$
\end{lemma}

\subsection{Continuity of entropies in fidelities}

Here we will establish a quantitative statement of the
continuity of the entropy as a function of the  density
operator, in terms of the fidelity of neighboring density
operators.

\begin{lemma} 
\label{lemma: entropy continuity}
For any density operators $\rho_1, \; \rho_2,$ acting on a
$d$-dimensional Hilbert space,
\beqa \label{eqtn: shmoo}
|S(\rho_1) - S(\rho_2)| \le 2 \sqrt{1 - F(\rho_1, \rho_2)} \log{d} 
+ 1
\eeqa
when
\beqa
2 \sqrt{1 - F(\rho_1, \rho_2)} < \frac{1}{3}\;.
\eeqa
\end{lemma}

\begin{proof}
The proof begins with an
inequality due to Fannes \cite{Ohya93a},
involving an ``error'' quantity different from $1-F(\rho_1,\rho_2)$.
Defining the $L_1$ norm of an operator $A$ as
\beqa
||A|| \equiv \tr |A| \equiv \tr \sqrt{A^\dagger A},
\eeqa
and the function $\eta(\cdot)$ by $\eta(x) = 
- x \log{x}$,
we have (when $||\rho_1 - \rho_2 || < \frac{1}{3}$)
\beqa \label{eqtn: L1normcontinuity}
|S(\rho_1) - S(\rho_2)| \le ||\rho_1 - \rho_2|| \log{d} 
+ \eta(||\rho_1 - \rho_2||)\;.
\eeqa
For our purposes, we may note that for $x < \frac{1}{3}$,
$\eta(x) < \frac{\log 3}{3} < 1$, and use the weaker inequality
\beqa \label{eqtn: L1normcontinuityweak}
|S(\rho_1) - S(\rho_2)| \le ||\rho_1 - \rho_2|| \log{d} + 1 \; .
\eeqa
Defining $p^{(1)}, p^{(2)}$ to be probability distributions
given by the eigenvalues of $\rho_1$ and $\rho_2$ respectively, 
we note that if the two density matrices
commute, then 
$||\rho_1 - \rho_2 || = 2 d_K(p^{(1)},p^{(2)})$, 
where $d_K$ is the {\em Kolmogorov distance} or
{\em total variation distance} between two probability
distributions,
\beqa
d_K(p^{(1)},p^{(2)}) \equiv 
\frac{1}{2} \sum_i |p^{(1)}_i-p^{(2)}_i|\;.
\eeqa
Since the entropy difference is invariant under independent
unitary rotations of each density matrix,
\beqa 
|S(\rho_1) - S(\rho_2)| \le 2 d_K(p^{(1)},p^{(2)})\log{d} + 1,
\eeqa
where we may take the eigenvalues to be arranged in order of
size in both probability distributions. 
An inequality of C. H. Kraft \cite{Kailath67a} \cite{Fuchs97c} implies
\beqa
d_K(p^{(1)},p^{(2)}) \le \sqrt{1 - B(p^{(1)},p^{(2)})}\;,
\eeqa
where $B$ is the Bhattacharyya-Wootters overlap
\beqa
B(p^{(1)},p^{(2)}) \equiv \sum_i \sqrt{p^{(1)}_ip^{(2)}_i}\;.
\eeqa
Moreover,
\beqa
B(p^{(1)},p^{(2)}) \ge F(\rho_1,\rho_2),
\eeqa
since, given the eigenvalues of both density operators, the fidelity
is maximized by choosing their eigenvectors to be the same,
assigned to eigenvalues in order of size.   This follows easily
from \cite{vonNeumann37a},\cite{Horn85a} and the representation of 
the square root of the fidelity as 
\beqa
\max_{{\rm unitary}\; U} \tr \rho_1^{1/2} \rho_2^{1/2} U\;.
\eeqa  
This completes the proof of the lemma.
\end{proof}

Now consider the situation where 
$d$ is the dimension of each of 
two spaces $Q$ and $R$, and $\rho^{RQ}$ a density operator on
the $d^2$--dimensional space $R \otimes Q$.  We use the notation
$\rho^Q \equiv {\rm tr}_R~\rho^{RQ}$.  
Using Lemma \ref{lemma: entropy continuity} and the fact that 
$F(\rho_1^Q,\rho_2^Q) \ge F(\rho^{RQ}_1, \rho^{RQ}_2)$, 
one easily obtains
a continuity relation for the entropy of 
$Q$ conditional on $R$, defined as
\beqa
S(Q|R) \equiv S(\rho^{RQ}) - S(\rho^Q)\;.
\eeqa

\begin{lemma}
\label{lemma: conditional entropy continuity}
Continuity of conditional entropy.
\beqa
|S(Q_1|R_1) - S(Q_2|R_2)| \le 6 \sqrt{1 - F(\rho^{RQ}_1, \rho^{RQ}_2)}
\log{d} + 2 \nonumber
\eeqa
when $F(\rho^{RQ}_1,\rho^{RQ}_2) > 5/9$.
\end{lemma}

This lemma will be useful in the discussion of capacity below, because
the capacity is bounded by a quantity, the {\em coherent information} of
a density operator $\rho^Q$ under an operation $\ce$, which may be written in terms of 
a conditional entropy.  This quantity is defined by
\beqa
I_c(\rho^Q, \ce) \equiv S \left(\frac{\ce(\rho^Q)}{\tr \ce(\rho^Q)}\right)
- S\left( \frac{\ci \otimes   \ce( |\psi^{RQ}\rangle \langle \psi^{RQ}|)}
{\tr \ce(\rho)}\right) \;.
\eeqa

\section{The typical subspace and entanglement fidelity}
We now derive some interesting implications of the 
QAEP for entanglement fidelity.
These may be summarized by the statement that
in order for the entanglement fidelity to be asymptotically
high, it is necessary and sufficient that the fidelity be high 
on the typical subspace. We will demonstrate two versions of
this statement, both of which will be used later on.

Define a {quantum data
compression scheme} for a source $\Sigma = (H_s,{\rho_s\pn})$ to be a sequence of  
trace-preserving quantum operations
${\cal C}\pn$ from $H_s^{\otimes n}$ to the $\epsilon$-typical subspace
$T\pn_\epsilon$ of the source 
such that 
\begin{eqnarray}
{\cal C}\pn(\rho_s\pn) =  {\cal C}_1\pn(\rho) + {\cal C}\pn_2(\rho_s\pn),
\end{eqnarray}
where ${\cal C}_1\pn(\rho) \equiv \Lambda_n \rho_s\pn \Lambda_n, 
\Lambda_n$ is the projector onto $T\pn_\epsilon$,
and $\tr (\Lambda_n \rho_s\pn) = 1 - \delta_n$.
Then we may derive the following lemma.
\begin{lemma}
For any source satisfying the QAEP, 
any quantum data compression scheme $\cc\pn$
and any trace-preserving operation ${\cal A}\pn$
from $H_s^{\otimes n}$ to
$H_s^{\otimes n}$,   
\begin{eqnarray}
|F_e(\rho_s\pn, {\cal A}\pn  {\cal C}\pn) - F_e(\rho_s\pn,{\cal A}\pn)| < 2 \delta_n.
\end{eqnarray}
\end{lemma}
This lemma is an immediate consequence of Lemma \ref{lemma: composition}.

In applying this lemma, we have in mind a situation where 
${\cal A}\pn$ represents the 
effect of further encoding taking
us from the source Hilbert space $H_s\on$
to the 
channel Hilbert space $H_c\on$, followed
by the channel noise operation, and a decoding which takes us 
back to the source Hilbert space.
By the  
QAEP, for large $n$ 
${\rm tr} \overline\Lambda\pn \rho_s\pn \overline\Lambda\pn \equiv
  \delta_n$ becomes smaller than any predetermined positive
$\delta$;  hence the difference between the entanglement 
fidelity when the encoding is preceded by quantum data compression
of the source, and the entanglement fidelity without such a 
step, is asymptotically negligible.

For some purposes, it will be more useful to compare the
entanglement fidelity of a source with the entanglement
fidelity of the {\em renormalized} projection of the
source onto its typical subspace.
Let $\Lambda$ be the projector onto
the typical subspace after $n$ uses of the source, and
$\overline \Lambda$ the projector onto the
orthogonal subspace.
For any positive $\epsilon$ and large enough $n$,
\begin{eqnarray}
\mbox{tr}(\overline \Lambda \rho_s\pn \overline \Lambda )
	\leq \epsilon.
\end{eqnarray}
Defining the renormalized 
restriction of the source to the typical subspace,
\begin{eqnarray}
\rho\pn_\epsilon \equiv \frac{\Lambda \rho_s\pn \Lambda}{
	\mbox{tr}(\Lambda \rho_s\pn \Lambda)},
\end{eqnarray}
and applying the continuity lemma for entanglement fidelity,
(\ref{lemma: continuity of entanglement fidelity}), 
we have the following lemma:
\begin{lemma} \label{lemma: renormalized typical fidelity}
For any trace-preserving operation ${\cal E}$ and any 
source satisfying the QAEP,
\begin{eqnarray}
|F_e(\rho_\epsilon\pn,{\cal E})-F_e(\rho_s\pn,{\cal E}) |
\leq
\frac{4 \epsilon}{(1-\epsilon)^2}\;. 
\end{eqnarray}
\end{lemma}
By choosing $n$ sufficiently large, $\epsilon$ can be
made arbitrarily small, and thus we see that
for the entanglement fidelity for the source to be
high asymptotically, it is necessary and sufficient that the
entanglement fidelity be high asymptotically for
the renormalized restriction of the source to the typical
subspace.

\section{Entanglement fidelity and minimum pure state fidelity}
\label{sec: equivalence}
\subsection{Entanglement transmission implies pure-state transmission}
We will first show that
if a source satisfying the QAEP can be transmitted over a channel with entanglement
fidelity approaching one in the large-block limit, 
one can transmit a subspace which is asymptotically of dimension
$2^{nS(\Sigma)}$ with minimum pure state fidelity approaching
one.  That is, if a channel can send entanglement at a certain
rate, it can send subspaces with high pure-state fidelity at 
that rate also.

The argument
has parallels with the classical argument that if one can 
transmit with low expected error (taking the expectation over
messages), one can transmit with low maximal error.  That
argument proceeds by throwing out the highest-error half of 
the codewords, and then establishing a definite bound on the
maximum error of the remaining codewords in terms of the 
average error of the initial ensemble of codewords.  Both
quantities go to zero together.  Throwing out half the codewords
reduces the rate by a bit, but asymptotically this is negligible.
Here, we throw out a low-fidelity fraction of the Hilbert space
dimensions, in a certain systematic way which enables us to
bound the minimum 
fidelity of the remaining states in terms of the entanglement
fidelity.

We do not expect to be able to show that a ``logarithmically
large'' subspace of the 
support of an arbitrary density operator may be sent with 
high minimum pure-state fidelity.  That would mean that the
capacity for sending subspaces with high minimum fidelity would
be higher than the capacity for sending entanglement, since the
dimension of the support of a density matrix is typically much
higher than its entropy.  After all, many of the dimensions
in the support of a density matrix may have negligible probability,
and hence the failure to send them would be expected to have 
negligible impact on the entanglement fidelity.  Therefore,
a high 
entanglement fidelity would not necessarily suggest that 
all dimensions
in the support, or even a logarithmically large subset of them,
can be sent accurately.  Rather, we expect to be able to show
that a subspace whose dimension is
a ``logarithmically large'' fraction of
$2^{nS(\Sigma)}$ can be sent with high minimum pure-state fidelity.
In fact, we expect that a logarithmically large subspace
of the typical subspace can be sent with high pure-state fidelity.

Our approach, then, will be to use Lemma \ref{lemma: 
renormalized typical fidelity} to argue
that if a source $\Sigma$ generating density operators
$\rho\pn$ can be sent with asymptotically high entanglement
fidelity, so can $\rho\pn_\epsilon$, the renormalized restriction of 
$\rho\pn$ to its $\epsilon$-typical subspace.  
Therefore, for large enough $n$, $\rho\pn_\epsilon$ can be 
made to  have entanglement fidelity greater than $1 - \eta$ for
any positive $\eta.$  We will indicate, without explicitly
changing notation from $\rho$ to $\rho_\epsilon\pn$, 
where we first use properties
of $\rho_\epsilon\pn$.  

Suppose a density operator $\rho$ with $K$-dimensional
support can be sent with entanglement
fidelity $1 - \eta$.  
Consider the following procedure for systematically removing
dimensions from the support of the density operator.
Let $|1\rangle$ be the lowest 
fidelity pure state in the support.  We then define the (sub-normalized) positive
operator $\rho_1$ by
\beqa
\tilde{\rho}_1 = \rho - q_1 |1\rangle \langle 1|\;,
\eeqa
where $q_1$ is the largest positive $q_1$ for which
$\tilde{\rho}$ is still a positive operator.  
We continue this process recursively, 
defining $\tilde{\rho}_0 \equiv \rho$, and
\beqa
\tilde{\rho}_{i} \equiv 
\tilde{\rho}_{i-1} - q_i |i\rangle \langle i|\;,
\eeqa
where $|i\rangle$ is the state in the support of $\tilde{\rho}_{i-1}$ 
with the lowest pure-state fidelity, and 
$q_i$ is as large as it can be subject to the constraint
that $\tilde{\rho}_{i}$ is a positive operator.

The vectors in this set are ordered in terms of 
increasing pure-state fidelity;  we will write 
$f_i$ for the pure state fidelity 
$\langle i | \ce(|i\rangle \langle i|)|i\rangle$ 
of $|i\rangle.$  

Note that $\tr \tilde{\rho}_1 = 1 - q_1$, and in general
$\tr \tilde{\rho}_j = \tr \tilde{\rho}_{j-1}
- q_j = 1 - \sum_{i=1}^j q_j$.  By construction, 
${\rm rank}(\rho_i) = {\rm rank}(\rho_{i-1}) - 1$.
Hence $\rho_d = {\bf 0}$ and $\sum_{i=1}^K q_i = 1$.
Furthermore,
\beqa
\sum_{i=1}^K q_i |i\rangle \langle i | = \rho\;,
\eeqa
that is, $\{q_i, |i\rangle\}$ are a pure-state
ensemble for $\rho$.    Note that while 
this procedure removes dimensions from the support 
of the density matrix one by one, the dimensions it 
removes are not necessarily the one-dimensional spaces
spanned by the vectors $|i\rangle$.  Indeed, the vectors
$|i\rangle$ 
will usually not be an orthonormal basis for the support
of $\rho$
although they are linearly independent.

Now 
\beqa
\la i | \rho | i \ra = q_i + \sum_{j \ne i} \la j | \rho | j \rangle.
\eeqa
Since the terms in the sum are all positive,
\beqa
q_i \le \la i | \rho | i \ra \le \lambda_1(\rho),
\eeqa
where $\lambda_1(\rho)$ is the largest eigenvalue of $\rho$.
That is, 
any upper bound on the eigenvalues of $\rho$ is also an upper bound on
the $q_i$.
 
In particular, when $\rho = \rho_\epsilon\pn$ then for large enough
$n$ the $q_i$ satisfy the bounds on eigenvalues from the QAEP
\beqa
2^{-n(S(\Sigma) + \epsilon)} \le q_i \le \frac{2^{-n(S(\Sigma) - \epsilon)}}
{1 - \delta} \;.
\eeqa

Now, by the convexity of entanglement fidelity in the density
operator,
\beqa
\sum_{i=1}^{n_0} q_i f_i + \bigl(\sum_{i = n_0 + 1}^{n} q_i \bigr)
F_e(\rho_{n_0+1}) \ge F_e(\rho) = 1 - \eta\;,
\eeqa
where $\rho_{n_0+1}$ is the normalized version of 
$\tilde{\rho}_{n_0+1}$, {\em i.e.}, the density operator
with the lowest-fidelity
$n_0$ dimensions of its support removed.
Define $\alpha \equiv \sum_{i = 1}^{n_0} q_i\;.$
Thus we are considering the
situation where we throw out $n_0$ of the states,
leaving a fraction $(1 - \alpha)$ of the total weight of the
density operator.

We will denote by $1 - \gamma$ the 
pure state fidelity of $|n_0 + 1\rangle$,
\beqa
f_{n_0+1} \equiv 1 - \gamma\; ;
\eeqa  this is the lowest
pure-state fidelity of any of the remaining vectors
$|i\ra$ for $i \ge n_0$, and
by construction also the lowest pure-state fidelity of any
state in the subspace they span.  
Then
\beqa
(1 - \gamma) \alpha + 
(1-\alpha) \ge 1 - \epsilon\;,
\eeqa
so that
\beqa
\gamma \le \frac{\eta}{\alpha}.  
\eeqa
Thus the reciprocal
of $\alpha$ is the factor by which error is increased when
the first $n_0$ dimensions are removed from the support of $\rho$ by
the above procedure.  
Since 
\beqa
2^{-n(S(\Sigma) + \epsilon)} \le q_i \;,
\eeqa
\beqa
n_0 2^{-n(S(\Sigma) + \epsilon)} \le \alpha\;.
\eeqa
Thus, for a fixed $\alpha$, our procedure leaves us with
a subspace having dimensionality $D \equiv K - n_0$ of pure states
which can be sent with fidelity at least $1 - \eta/\alpha$.
Now,
\beqa
1 - \alpha = \sum_{i=n_0+1}^K q_i \le D 2^{-n(S - \epsilon)}
\eeqa
so
\beqa
D \ge (1 - \alpha) 2^{n(S - \epsilon)}
\eeqa
and the rate 
\beqa
\frac{\log{D}}{n} \ge \frac{\log{(1 - \alpha)}}{n} + S(\Sigma) - \epsilon\;.
\eeqa

That is, for any fixed $\eta$ and $\alpha$ strictly between zero and 
one, for large enough $n$ the size in qubits of a
subspace with minimum fidelity 
$1- \eta / \alpha$ approaches $n(S(\Sigma ) - \epsilon)$. 
Hence
all rates less than $S(\Sigma)$ may be achieved.  Since $\Sigma$ was any
density operator source that could be sent with high entanglement fidelity,
this implies that any rate less than the capacity for sending entanglement 
may be achieved for sending subspaces with high minimum
entanglement fidelity.  Thus we have
\begin{theorem}
$Q_s \ge Q_e$.
\end{theorem}

\subsection{Pure state transmission implies entanglement transmission}

We now show that the entanglement fidelity 
of a density operator under an operation
cannot be too much less than the 
minimum pure-state fidelity of states in the density 
operator's support.  As minimum pure-state fidelity approaches
one, so does entanglement fidelity, so that {\em any} 
density operator with support entirely in this subspace 
can be sent with high entanglement fidelity. 
Specifically, we prove the following theorem (see also \cite{Knill97a}).
The argument makes no use of the notion of typical subspace, and hence is
not limited to sources satisfying the QAEP.
\begin{theorem}
Suppose all pure states $|\psi\rangle$ in a 
subspace $S$ have pure state fidelity 
$\langle \psi |\ce(|\psi \rangle \langle \psi |)|\psi \rangle$
greater than or
equal to $1 - \eta$. 
Then any density operator $\rho$ whose
support lies entirely in that subspace 
has entanglement fidelity $F_e(\rho,\ce) \ge 1 - \frac{3}{2}\eta$.
\end{theorem}
For applications to asymptotic channel 
capacity  what is
important is that the error for sending entanglement goes to zero
if the maximum error for density operators in the subspace does,
and that the relationship between the two fidelities involves no
factors of the dimension of Hilbert space, which could cause
trouble in taking the large block limit.
This means that if we can transmit Hilbert-space dimensions
with minimum fidelity approaching one at a rate $C$, we can
also reliably transmit the entanglement of 
any source $\Sigma$ with entropy $S(\Sigma) < C$. 

\begin{proof}
We Schmidt decompose $|\Psi^{RQ}\rangle$:
\beqa
|\Psi^{RQ}\rangle = \sum_k \sqrt{\lambda_k}|k^R\rangle |k^Q\rangle\;.
\eeqa
In the Schmidt decomposition
$|k^R\rangle$ and $|k^Q\rangle$ are the diagonal bases
of the density operators $\rho^R$ and $\rho^Q$, labeled
according to their common eigenvalues $\lambda_k$.

Then
\beqa
\rho^{RQ'} & \equiv & (\ci \otimes \ce)(|\psi^{RQ}\rangle \langle
\psi^{RQ}|) \nonumber \\
& = & \sum_{kl} |k^R \rl l^R| \otimes \ce (|k^Q \rl l^Q|)\;.
\eeqa
The entanglement fidelity becomes (omitting the superscripts
$R$ and $Q$ to reduce clutter):
\beqa \label{eqtn: entanglement fidelity in Schmidt decomp}
F_e(\rho, \ce) & = &\sum_{mnkl} 
\sqrt{\lambda_m \lambda_n \lambda_k \lambda_l} 
\langle m|n \rl l | n \rl m| \ce(|k \rl l|)|n\rangle  \nonumber \\
& = & \sum_{k l} \lambda_k \lambda_l 
\langle k| \ce(|k \rl l|)|l\rangle \;.
\eeqa
A first attempt at a proof might split up the sum as: 
\beqa
F_e & = & \sum_k \lambda_k^2 \langle k| \ce(|k \rl k|)|k\rangle \nonumber \\
& + & \sum_{k \ne l} \lambda_k \lambda_l 
\langle k| \ce(|k \rl l|)|l\rangle \;.
\eeqa
We see that the first sum here can certainly be bounded below 
using the fact that pure state fidelities for vectors in the basis 
$|k\rangle$ are greater than 
$ 1 - \eta$,  but the second term has cross-terms that are
more difficult to deal with.  The proof will have to 
use the fact that
not only vectors in the basis $|k\rangle$, but arbitrary 
superpositions of them, have high  fidelity, and the pure
state fidelities of these superpositions will contain such 
cross-terms.  Since the expressions we want to bound contain
the probabilities $\lambda$, we will consider superpositions
with amplitudes $\sqrt{\lambda}$
and all possible phase factors $e^{i \phi_k}$:
\beqa
|\psi(\phi_1,...,\phi_k)\rangle
\equiv 
\sum_k \sqrt{\lambda_k}e^{i \phi_k}|k \rangle\;,
\eeqa

The pure state fidelity for this is:
\beqa \label{eqtn: dingbat}
&&\langle \psi| \ce(|\psi \rl \psi|)|\psi\rangle
\nonumber \\
& = & \sum_{mnkl}
\sqrt{\lambda_m \lambda_n \lambda_k \lambda_l}
\langle m| \ce(|k \rl l|)|n\rangle 
e^{i(\phi_k + \phi_n - \phi_m -\phi_l)}\;. \nonumber \\
{}
\eeqa

The $m=k, n=l$ terms 
will give the entanglement fidelity in the form
(\ref{eqtn: entanglement fidelity in Schmidt decomp})
(since the phases appear in complex conjugate pairs in
those terms, they disappear).
But there are other terms in (\ref{eqtn: dingbat})
which we need to argue are small, or somehow
get rid of, in order to argue that the high fidelity
of these pure states implies that the terms constituting
the entanglement fidelity are high.
We do this by averaging the entanglement fidelity for these superpositions
over all phases from zero to $2 \pi$.  We still get the
desired terms, but many of the cross terms will disappear.  Only
those with four indices identical, or with indices identical in
complex conjugate pairs, will remain;  the rest will contain
integrals like $\int_0^{2\pi}d\phi_k  e^{i\phi_k}$ (from indices 
whose value is not equal to that of some other index), or 
$\int_0^{2\pi}d\phi_k  e^{2i\phi_k}$ (from pairs of identical
indices that are not complex conjugates).  The average is of 
course still greater than $1-\eta$;  and the remaining terms
are:
\beqa
\overline{f} &=& \sum_{k l} \lambda_k \lambda_l 
\langle k| \ce(|k \rl l|)|l\rangle 
+ \sum_{km,k\ne m} \langle m|\ce(|k \rl k|)|m\rangle \nonumber \\
&=& F_e + 
\sum_{km,k\ne m} \langle m|\ce(|k \rl k|)|m\rangle
\ge 1 - \eta\;.
\eeqa
(The same result obtains if the average is done not by 
integration over all possible values of the complex
phase factor, but only over the phases $\pm 1, \pm i$.)
We need to upper bound the terms that do not appear in the
entanglement fidelity.  These terms all contain the fidelity
of the output state $\ce(|k \rl k|)$
to a state $|m\rangle$ orthogonal to the input state $|k\rangle$.
Since $\ce(|k\rl k|)$ has high fidelity to input $k$, 
one expects
its fidelity to states orthogonal to $|k\rangle$ will be small.
In fact, since $\ce$ is trace-preserving, taking the trace
in the $|m\rangle$ basis gives:
\beqa \label{eqtn: sum bound from trace-preserving}
\sum_m \langle m|\ce(|k \rl k|)| m\rangle = 1 \;,
\eeqa
and the fact that $\langle k| \ce(|k \rl k|)|k\rangle 
\ge 1 - \eta$ then gives:
\beqa
\sum_{m \ne k}  \langle m|\ce(|k \rl k|)| m\rangle \le \eta \;.
\eeqa
Let us assume the eigenvalues have been ordered from largest 
to
smallest, 
$\lambda_1 \equiv \lambda_1(\rho) 
\ge \lambda_2 \equiv \lambda_2(\rho)$
etc.
Then when $k = 1$, we have $\lambda_m \le \lambda_2$, so the
$k=1$ term is bounded:
\beqa
\lambda_1 \sum_{m \ne 1} \lambda_m \langle m|\ce(|1 \rl 1|)
| m\rangle &\le& \lambda_1 \lambda_2 
\sum_{m \ne 1} \langle m|\ce(|1 \rl 1|)| m\rangle \nonumber \\
&\le& 
\lambda_1 \lambda_2 \eta \;.
\eeqa
When $k \ne 1$, we must use the looser bound $\lambda_m 
\le \lambda_1$ in a similar fashion, giving:
\beqa
& &\sum_{k \ne 1}
\lambda_k \sum_{m \ne k} \lambda_m \langle m|\ce(|k \rl k|)
| m\rangle \nonumber \\
& \le & \sum_{k \ne 1} \lambda_k \lambda_1
\sum_{m \ne k} \langle m|\ce(|k \rl k|)| m\rangle \nonumber \\
& \le & 
\sum_k \lambda_k \lambda_1 \eta = (1 - \lambda_1)\lambda_1 \eta \;.
\eeqa
Thus
\beqa
F_e \ge 1 - (1 + \lambda_1 \lambda_2 + 
(1 - \lambda_1)\lambda_1)\eta\;.
\eeqa
For given $\lambda_1$, this is minimized where $\lambda_2 = 
(1 - \lambda_1)$.  The resulting bound,
\beqa
F_e \ge 1 - (1 + 2 \lambda_1 (1 - \lambda_1))\eta \;,
\eeqa
is clearly minimized when $\lambda_1 = \lambda_2= \frac{1}{2}$, 
giving
\beqa
F_e(\rho,\ce) \ge 1 - \frac{3}{2}\eta\;.
\eeqa
\end{proof}

\begin{corollary}
$Q_e \ge Q_s$.  
\end{corollary}
\begin{proof}
The theorem implies, as noted in \cite{Barnum97a}, that
if there is a sequence of encodings,
decodings, and subspaces $H\pn$ 
that achieves rate $R$ for subspace transmission, the sequence
of uniform
density operators on these subspaces  $I\pn/{\rm dim}(H\pn)$ will also have  
limiting entanglement fidelity one under the same transmission operations.
Since the entropy rate of this source is $R$, the same rate is achievable
for entanglement transmission.
\end{proof}

\subsection{Consequences for Capacity}
The results of the two previous sections immediately
imply that the capacities
for pure-state transmission and for entanglement transmission are 
equal.
They also imply 
that if a source can be sent on a given channel
with high entanglement 
fidelity, so can any source with lower entropy which satisfies
the QAEP.
Hence
\begin{theorem}
Any source $\Sigma$ with $S(\Sigma) < C(\cn)$ may be transmitted
with high entanglement fidelity over the channel $\cn$.
\end{theorem}

\section{Encodings}

\thispagestyle{empty}

In \cite{Barnum97a}, 
we conjectured that an expression for the quantum capacity was:
\beq
\lim_{n \rightarrow \infty} \max_{H_s, \rho\pn \in H_s, \ce\pn: B(H_s) \rightarrow B(H_c)} 
I_c(\rho\pn,
\cn\on \ce\pn) \;.
\eeq
and showed that this expression was no smaller than the
capacity.
This involves a maximization over input density operators and
trace--preserving
completely positive
encoding maps.  However, we also conjectured that the maximization over
encodings was not necessary.  Rather, the analogous
expression
with the maximization over encodings removed, and the density 
operator maximization done over density operators on the channel input
Hilbert
space $H_c$, instead of operators on the source space, was conjectured to
also be a correct
expression for the capacity.  This would make the situation
more similar to the classical one, where no maximization over encodings 
appears in the expression for channel capacity.  In \cite{Barnum97a}, we
showed that if encodings could be restricted to be unitary, then indeed
the maximization over encodings could be dropped entirely.  

\subsection{Partially isometric encodings}
\label{subsec: partially isometric encodings}
Our  strategy for removing the maximization over encodings will be to show
that we may restrict our attention to {\em partially isometric}
encodings, that is,
encodings of the form 
\beqa
\ce(\rho) = V \rho V^\dagger,
\eeqa
where $V$ is a partial isometry from the source space to the 
channel space. 
An encoding corresponding to a partial 
isometry from a source space to a smaller
channel space
(as in noiseless 
data compression, for instance) will be trace-decreasing
for density operators having support outside the subspace that is 
unitarily mapped into the source space. 
In our definition of channel capacity, we required that encodings be
trace-preserving.
But trace-decreasing encodings are relevant to our problem because  
they may be embedded in trace-preserving ones with no loss of fidelity.
We say a trace-decreasing operation
$\cf$ is {\em embedded} in a trace-preserving operation $\ca$ if
\beqa
\ca = \cf + {\cal G}
\eeqa
for some trace-decreasing ${\cal G}$.
Since 
\beqa
F_e(\rho,\cf + \cg) = F_e(\rho, \cf) + F_e(\rho, \cg)\;,
\eeqa
the entanglement
fidelity of a trace-decreasing operation is a 
lower bound on the entanglement fidelity of any trace-preserving
operation into which the trace-decreasing one has been embedded.
This is what makes partially isometric encodings relevant to physical
situations in which a trace-preserving encoding is used; we will
use this in \ref{sec: upperbound}.

\section{Restricting the encodings}

We will show that if 
there exists a general encoding
that achieves high fidelity transmission for a given source, 
there is also a partially
isometric encoding achieving
fidelity not much lower for that source, 
where ``not much lower'' will be quantified in 
such a way that if the general 
encoding has fidelity approaching one, the lower bound on 
fidelity with partially isometric encoding also approaches
one, and there is no dimensional 
dependence in the relation between the
fidelities that would cause difficulty with the large block limit (in
which the Hilbert space dimension grows exponentially).

\subsection{Perfect transmission}
The intuition behind the argument may be illustrated for the case of 
transmission with fidelity precisely one.  
It is frequently easy to show something
for fidelity exactly one, but more 
difficult to extend it to
fidelities which are merely very close to one, as is necessary for 
channel capacity arguments, and that is the case here.  If the 
operation of encoding followed by noise followed by decoding achieves
perfect transmission for some $\rho$, 
this implies that the encoding operation is
perfectly reversible for $\rho,$ since it is reversed by the composition
of noise with decoding.  As noted
in \cite{Nielsen97a}, an operation $\ca$ that is perfectly 
reversible for a density operator may, when restricted to the subspace $C$
(with dimension $d_C$)
supporting that density operator,
be written in the form of unitaries 
from the support into mutually orthogonal $d_C$-dimensional 
subspaces of the output space,
randomly applied with probabilities $p_i$.  That is
\beqa
\ca \sim \{\sqrt{p_i} U_i\}\;, \nonumber \\
U_i^\dagger U_j = \delta_{ij} P_C 
\eeqa
where $P_C$ is the projector onto $C$.
If there exists an operation which reverses this with perfect
fidelity for some input $\rho,$ it must reverse each of the unitaries
$U_i$ with fidelity
one.  Hence we may remove the factor $\sqrt{p_i}$ from any of the
operators in the canonical decomposition of the encoding, 
and use it as
an encoding, which will achieve perfect transmission when the same 
decoding is used.

\subsection{Isometric encoding suffices}

\begin{theorem} \label{theorem: isometric}
Given a trace-preserving map $\ca$ and a map $\ce$ with 
$\tr \ce(\rho) = 1$ and 
\beqa
F_e(\rho, \ca    \ce ) > 1 - \eta\;,
\eeqa
there exists a partial isometry $W$ such that
\beqa
F_e (\rho, \ca  W ) > 1 - 2 \eta \;.
\eeqa
\end{theorem}

In applying this theorem, we will take $\ce$ to be the encoding map,
and $\ca$ to be the concatenation of noise and decoding.

The proof proceeds via the following two lemmas:

\begin{lemma} \label{lemma: single composition}
There exist operator decompositions
of $\ca$ and $\ce$ such that $F_e(\rho,\ca\ce)\leq F_e(\rho,
A_1 E_1/\sqrt{\tr(E_1\rho E_1^\dagger)})$.
\end{lemma}

(Note that $E_1/\sqrt{\tr(E_1\rho E_1^\dagger)}$ is not necessarily trace
decreasing.)

\begin{proof}
Let $\{A_i\}$ and $\{E_i\}$ be operator decompositions of
$\ca$ and $\ce$. Let $X$ be the matrix with elements
$\tr(A_iE_j\rho)$.  Then $F_e = \sum_{ij} |(X)_{ij}|^2$.
The singular value decomposition ensures that by changing the operator
decomposition of $\ca$ and $\ce$, we can transform to a representation
where $X$ is diagonal; assume without loss
of generality that $A_i$ and $E_j$ are already such representations.
Then $F_e(\rho,\ca\ce) = \sum_k\tr(A_kE_k\rho)^2$
(since $\tr(A_kE_j\rho) = 0 $ if $k\not=j$).
Let $\lambda_k = \tr(E_k \rho E_k^\dagger)$.
Then $\sum_k \lambda_k (\tr(A_kE_k\rho)^2/\lambda_k) = F_e(\rho,\ca\ce)$
and $\sum_k\lambda_k = 1$,
so there exists a $k$ such that
$\tr(A_kE_k\rho)^2/\lambda_k \geq F_e(\rho,\ca\ce)$.
\end{proof}

\begin{lemma} \label{lemma: there is a unitary}
Let 
\beqa
E&:& S \rightarrow C \nonumber \\
A&:&C \rightarrow S
\eeqa
be linear operators, $\rho \in B(C)$ a 
density matrix.
If $F_e(\rho, AE)\geq 1-\eta$, $A^\dagger A\leq I$
and $\tr(E\rho E^\dagger)=1$, then there is
a maximal partial isometry $W: S \rightarrow C$ 
such that $F_e(\rho, AW)\geq 1-2\eta$.
\end{lemma}

\begin{proof}
Let $UD_AV$ be a singular value decomposition of $A$. 
Here we can take $D_A$ to have matrix elements
proportional to the Kronecker 
delta in a (fixed) basis for $C$ , $V$ unitary on 
$C$ and $U: C \rightarrow S$ a maximal partial isometry.
Consider the maximal partial isometry $W: S \rightarrow C$ 
defined by $W = V^\dagger U^\dagger$.
Then $U = (VW)^\dagger$ and 
\begin{eqnarray}
|\tr(AE\rho)|^2 &=&
    |\tr(\rho^{1/2}UD_A^{1/2}VWUD_A^{1/2}VE\rho^{1/2})|^2 \nonumber \\
  &\leq&
    \tr((UD_A^{1/2}VW)^\dagger \rho UD_A^{1/2}VW) \nonumber \\
    &{}& \times \tr(UD_A^{1/2}VE\rho E^\dagger (UD_A^{1/2}V)^\dagger)\\
  &\leq&
    \tr(UD_A^{1/2}VW(VW)^\dagger D_A^{1/2}U^\dagger\rho)
    \label{eq:less1}\\
  &=&
    \tr(UD_AU^\dagger\rho)\nonumber \\
  &=&
    \tr(UD_AVW\rho).
\end{eqnarray}
(The first inequality is an operator Schwarz inequality, while
the second is due to
the fact that $A^\dagger A$, and therefore
$D_A$, is less than or equal to $I$, and
the fact that if $B \ge 0$ and $I \ge C \ge 0$, 
$\tr BC \le \tr B$.)
It follows that
$\tr(AW\rho)\geq 1-\eta$, hence
$|\tr(AW\rho)|^2  = F_e(\rho,AW)\geq 1-2\eta$.
\end{proof}

To obtain Theorem \ref{theorem: isometric} as a corollary,
apply Lemma \ref{lemma: single composition} 
and the premise of the theorem to get:
\beqa
F_e(\rho,A_1 E_1/\sqrt{\tr(E_1\rho E_1^\dagger)}) \ge 
F_e(\rho, \ca \ce) \ge 1 - \eta\;.
\eeqa
Lemma \ref{lemma: there is a unitary}, with 
\beqa
E = E_1/\tr(E_1 \rho E_1^\dagger)\;
\eeqa
then gives the result.

It follows 
that if there exists 
a coding scheme which transmits the 
entanglement of a source
reliably, there exists a
coding scheme using partially isometric encodings which transmits
it reliably.

\section{Forward classical communication doesn't help}
Bennett, DiVincenzo, Smolin and Wootters (BDSW) \cite{Bennett96a}
showed that a forward classical channel, from encoder to 
decoder, cannot help one achieve perfect transmission.
They did this by constructing, from any fidelity-one
coding scheme for such a channel, a fidelity-one coding
scheme which makes no use of the classical channel.  This
is another example of a result which is apparently hard
to extend to asymptotically high fidelity transmission. 

An argument virtually identical to the proof of Theorem
\ref{theorem: isometric} can be used to extend their
result to the asymptotically high fidelity situation, for
the problem of sending the entanglement of a uniform source
with asymptotically high fidelity (cf. also \cite{Barnum97b}).  
(By results in Section
(\ref{sec: equivalence}), this is 
equivalent to the problem BDSW considered, of 
sending every state in a source space with high 
fidelity.)  We may do this by modeling a classical forward 
channel in a manner analogous to the model of the observed
channel in \cite{Barnum97a}
except that the decoder takes into
account classical information about the encoding rather than
the noise.  We take the encoding to be
a set $\{ \ce_m \}$ of trace-nonincreasing operations
which sum to a trace-preserving operation.  The value of the
index $m$ represents classical information available to the
encoder (as a measurement result, say) which may be sent to
the decoder and used in decoding, so we allow the decoder to
use one of a collection of trace-preserving decodings $\cd_m$.
Formally, we define a {\em coding scheme with classical forward
communication} to be a sequence of such collections of encodings
and decodings, $[ \ce\pn_{m},\cd\pn_{m} ]$, where $m$ takes
values $1...M\pn$ so that the number of available encoding 
operations may depend
on $n$, and
\beqa
\sum_{m=1}^{M\pn} \ce\pn_m \equiv \ce\pn
\eeqa
is trace-preserving, while each of $\ce\pn_m$ is trace-nonincreasing.
We say a source may sent reliably over this channel with classical
forward communication, if there exists a coding scheme such that
\beqa
\lim_{n \rightarrow \infty} 
\sum_m F_e(\rho\pn, \cd\pn_m   \cn\on   \ce\pn_m) = 1\;.
\eeqa
We define the {\em capacity of a channel
for entanglement transmission with
forward classical communication}, $Q_e^{(fc)},$ to be the supremum
of the entropy rates of sources 
that can be sent reliably on the channel.

Now suppose the entanglement fidelity of a density operator $\rho$
sent through such a channel is high (omit the superscripts $\pn$ for
clarity), 
\beqa \label{eqtn: classicalforwardfidelity}
\sum_m F_e(\rho, \cd_m   \cn   \ce_m) > 1 - \eta.
\eeqa
Then there exists a value of $j$ of the index $m$ for which
\beqa
\hat{F}_e(\rho, \cd_j \cn  \ce_j) 
> 1 -  \eta.
\eeqa
Now
\beqa
\hat{F}_e(\rho, \cd_j \cn  \ce_j) = F_e(\rho, \cd_j \cn 
\frac{\ce_j}{\tr \ce_j(\rho)})\;,
\eeqa
so that by Theorem \ref{theorem: isometric}, there is a
partial isometry $W$ such that
\beqa
F_e(\rho, \cd_j \cn W) > 1 - 2 \eta\;.
\eeqa
The partially isometric encoding can 
be extended to a trace-preserving encoding with no loss of fidelity, 
hence the same source can be 
sent without using the forward classical channel.
Hence
\begin{theorem}
$Q_e^{(fc)} = Q_e\;.$

\end{theorem}
\section{An upper bound on capacity}\label{sec: upperbound}
We will now treat an issue raised in section
\ref{subsec: partially isometric encodings}.  We will show
that the fact that partially isometric encodings suffice to achieve
the channel capacity implies that we may omit the maximization over
encodings from the expression that upper bounds the capacity.

Since the entanglement fidelity of any trace-preserving encoding into which a
(possibly trace-decreasing) partially isometric encoding
 $\cv$ might be embedded is 
bounded below by the unrenormalized entanglement fidelity of $\cv$,
we consider trace-preserving encodings $\cf \equiv \cv + \ca$, where
$\cv$ is partially isometric. 
Polar decompose $V$ into
a maximal partial isometry $W$ and a positive $\Gamma$ (which will be a 
projector), so that $V = W\Gamma$.

We know from Theorem \ref{theorem: isometric} that, given a sequence of general encodings $\ce$  and decodings 
$\cd$ that
sends a given source (so that overall entanglement fidelity goes to one
with increasing block size), there exists a sequence of partially isometric 
encodings $\cv\pn$ 
that (when used with the same decodings as before), sends that 
source with the unrenormalized entanglement fidelity approaching one
with increasing block size.  Hence the entanglement fidelity when some 
sequence of trace-preserving extensions $\cf\pn= \cv\pn + \ca\pn$ 
is used to encode goes to one 
with increasing block size as well.  
More precisely, if for a given $\epsilon$ and large enough $n$, we have
\beqa
F_e(\rho\pn,\cd\pn   \cn\on  \ce\pn) > 1 - \epsilon
\eeqa
then by Theorem \ref{theorem: isometric}
for large enough $n$
\beqa
F_e(\rho\pn, \cd\pn    \cn\on   \cf\pn) > 1 - 2 \epsilon \;.
\eeqa

Now let us consider the fidelity of the output states $\rho^{RQ''}
\equiv \cd\pn    \cn\on   \ce\pn(\rho\pn)$ and
$\sigma^{RQ''}\equiv \cd\pn    \cn\on   \cf\pn(\rho\pn)$ obtained by
using the different encodings. By Lemma \ref{lemma: closefinal},
\beqa
F(\rho^{RQ''},\sigma^{RQ''}) > 1 - 3 \epsilon\;.
\eeqa

To obtain the upper bound on capacity in \cite{Barnum97a}, we used
the following fact:

\beqa
S(\rho) \le I_c(\rho\pn, \cn\on   \ce\pn) + 2 \nonumber \\ 
+ 4 (1 - F_e(\rho, \cd\pn  
\cn\on   \ce\pn))\log{d_c^n}\;,
\eeqa
where all operations involved are trace-preserving ($d_c$ is 
the dimension of the channel).
In particular, this holds for the operations $\cf\pn$.  We now
consider the coherent information with such encoding operations.

Recall the representation of the coherent information 
$I_c$ as a conditional entropy
and apply Lemma \ref{lemma: conditional entropy continuity}, 
the continuity of conditional entropy, to obtain:
\beqa
|I_c(\rho\pn, \cn\on   \ce\pn) &-& I_c(\rho\pn, \cn\on   \cf\pn)| \nonumber
\\
& < &  6 \sqrt{3 \epsilon}
\log{d_c^n} + 2\;.
\eeqa

Hence
\beqa
\lim_{n \rightarrow \infty} & &  
| \max_{\rho_1\pn}\frac{I_c(\rho_1\pn, \cn\on   \ce\pn)}{n} \nonumber \\
& - &  \max_{\rho_2\pn}\frac{I_c(\rho\pn, \cn\on   \cf\pn)}{n}
| = 0\;.
\eeqa
So, the coherent information bound with general encodings is the 
same as the bound for encodings restricted to have the form $\cf$.

We now show that this bound implies that with the maximization 
over channel input density operators alone.
In earlier work \cite{Barnum97a}, we defined the coherent information of a non-trace-preserving
operation as:
\beqa
I_c(\rho, \ce) \equiv S\left(\frac{\ce(\rho)}{\tr \ce(\rho)}\right)
- S\left( \frac{\ci \otimes   \ce( |\psi^{RQ}\rangle \langle \psi^{RQ}|)}
{\tr \ce(\rho)}\right) \;,
\eeqa
the conditional entropy using the {\em renormalized} output state of 
the system and entangled reference.
Now, the coherent information of the channel $\cf$ is bounded above by 
the coherent information of the {\em observed} channel $\cn  
\{\cv,\ca\}$
in which we know
which of $\cv$ and $\ca$ occurred.  The latter is given \cite{Barnum97a} by:
\beqa
I_c(\rho,  \cn\on  \{\cv,\ca\})
=
(\tr \Gamma \rho) I_c(\rho,  \cn\on \cv) \nonumber \\
+ (1 - \tr \Gamma \rho)I_c(\rho,  \cn\on   \ca) \;.
\eeqa
A straightforward calculation shows that the first term is equal to 
\beqa
\tr \Gamma \rho I_c(\widehat{W\Gamma \rho \Gamma W^\dagger},  \cn\on) \;.
\eeqa
(We use the notation $\hat{A} \equiv A/{\tr A}$.)
Since $1 \ge \tr \Gamma \rho \ge F_e(\rho,  \cd   \cn\on   \cv)$ and
the latter approaches one in the large-$n$ limit, so does $\tr \Gamma \rho$,
and hence:

\beqa\label{eq: a nice expression}
S(\Sigma) \le 
\lim_{n \rightarrow \infty} \frac{I_c(\widehat{W \Gamma \rho \Gamma W^\dagger}
,  \cn\on)}{n}\;.
\eeqa

The inequality still holds when we maximize over $\rho$. The ability
to maximize over $\rho$ followed by projection using $\Gamma$, 
normalization, and placing
the density operator in some subspace of the channel via $W$
just allows us to access some of the possible 
channel input density matrices.  Hence the RHS of 
(\ref{eq: a nice expression}) is bounded above by:
\beqa
\lim_{n \rightarrow \infty} \max_{\rho\pn}\frac{I_c(\rho\pn,  \cn\on)}{n}\;.
\eeqa
This is the promised upper bound on the quantum capacity.

\section*{Acknowledgments} 
H.B. and M. A. N. thank the Office of Naval Research for 
financial support under grant No. 
N00014-93-1-0116 while they were with the Center for Advanced
Studies, Dept. of Physics and Astronomy, University of New 
Mexico, and Carlton M. Caves for valuable discussions.   
H. B. thanks  the National Science Foundation 
for financial support under grant
PHY-9722614, and the Institute for Scientific Interchange Foundation, Turin, 
Italy, and ELSAG-Bailey for financial support.  E. K. thanks the National
Security Agency for support.

%\nocite{*}
%\bibliographystyle{IEEE}
%%%%%\bibliography{bib-file}  % commented if *.bbl file included, as
%%%%%see below
%\bibliography{bib}

%%%%%%%%%%%%%%%%% BIBLIOGRAPHY IN THE LaTeX file !!!!! %%%%%%%%%%%%%%%%%%%%%%%%
%% This is nothing else than the IEEEsample.bbl file that you would         
%%
%% obtain with BibTeX: you do not need to send around the *.bbl file        
%%
%%---------------------------------------------------------------------------%%
%
%
%%---------------------------------------------------------------------------%%

%\begin{biography}Howard Barnum is a swell dude.
%\end{biography}
%\begin{biography}
%Emmanuel Knill is an even sweller dude.
%\end{biography}
%\begin{biography}
%Michael Nielsen is the cat's pajamas.
%\end{biography}
\end{document}